\documentstyle[aps,prl,psfig]{revtex}
\title{\bf Diamond and $\beta$-tin structures of Si studied with
quantum Monte Carlo calculations}

\author{D. Alf\`{e}$^{1,2}$ and M. J. Gillan$^2$
\smallskip \\ $^1$Department of Earth Sciences, University College
London \\ Gower Street, London WC1E~6BT, UK
\smallskip \\ $^2$Department of Physics and Astronomy, University
College London \\ Gower Street, London WC1E~6BT, UK
\bigskip \\ M. D. Towler and R. J. Needs
\smallskip \\ Cavendish Laboratory, University of Cambridge, Cambridge
CB3 0HE, UK}

\begin{document}

\maketitle

\begin{abstract}
We have used diffusion quantum Monte Carlo (DMC) calculations to study
the pressure-induced phase transition from the diamond to $\beta$-tin
structure in silicon. The calculations employ the pseudopotential
technique and systematically improvable B-spline basis sets. We show
that in order to achieve a precision of 1 GPa in the transition
pressure the non-cancelling errors in the energies of the two
structures must be reduced to 30~meV/atom. Extensive tests on system
size errors, non-local pseudopotential errors, basis-set
incompleteness errors, and other sources of error, performed on
periodically repeated systems of up to 432 atoms, show that all these
errors together can be reduced to well below 30~meV/atom. The
calculated DMC transition pressure is about 3~-~4~GPa higher than the
accepted experimental range of values, and we argue that the
discrepancy may be due to the fixed-node error inherent in DMC
techniques.
\end{abstract}

\section{Introduction}

The importance of the quantum Monte Carlo technique (QMC) for
computing the energetics of condensed matter is becoming ever more
widely appreciated.  Even though its computational demands are much
greater than those of standard density functional theory (DFT), its
considerably greater accuracy for many systems~\cite{foulkes01} makes
the additional effort well worthwhile. Indeed, QMC is often seen as
one of the key ways of assessing the inadequacies of
DFT~\cite{filippi02,leung99,grossman95}. Nevertheless, QMC itself is
not exact, and it is important to probe its accuracy for different
kinds of problem. A sensitive way of doing this is to examine the
relative energies of different crystal structures of a material.  We
present here a QMC study of the energetics of the diamond and
$\beta$-tin structures of silicon; we calculate their total energies
as a function of volume, and hence the transition pressure between the
structures, for which there are experimental
data~\cite{experiments,mujica03}. We analyse in detail the sources of
the QMC errors, and use the comparison with experiment to gauge the
likely size of errors that cannot be eliminated.

The QMC calculations are performed within periodic boundary
conditions. Only the valence electrons are treated explicitly, the
interactions between valence and core electrons being represented by
pseudopotentials. We perform two type of QMC calculations: variational
Monte Carlo (VMC) and diffusion Monte Carlo (DMC). DMC results are
considerably more accurate, but VMC plays an indispensable role
because it provides the optimised trial many-electron wavefunctions
needed in DMC. This set of techniques is described in detail in a
recent review~\cite{foulkes01}, and implemented in the {\sc casino}
code~\cite{needs04} used in this work. The techniques are known to
give cohesive energies for group IV elements in the diamond structure
in very close agreement (within 100 meV/atom) with experimental
values. They have also indicated substantial DFT errors in, for example,
the formation energy of self-interstitials in Si~\cite{leung99}, 
the energetics of H$_2$
dissociation on Si (001)~\cite{filippi02}, and the energies of carbon
clusters~\cite{grossman95,kent00}.

The primary quantity calculated in this work is the total energy per
atom in the perfect crystal.  This energy is subject to different
kinds of error. The first kind consists of errors that can in
principle be reduced below any specified tolerance, for example
statistical error, time step and population control bias, basis-set
error in the trial wavefunction, and error due to the limited size of
the periodically repeated cell. Then there are errors that cannot be
systematically eliminated, but whose size can at least be estimated by
purely theoretical means. The main error of this kind comes from the
so called ``pseudopotential localisation approximation'', which cannot
be avoided in present QMC techniques based on non-local
pseudopotentials~\cite{nonlocal}. Finally, there are errors that
cannot be eliminated and are also difficult to assess except by
comparison with experiment.  There is only one error of this kind, the
so called QMC ``fixed-node error''. Our strategy in this work will be
to demonstrate that all errors of the first kind have been made
negligible, do our best to estimate errors of the second kind, and
then appeal to experiment to assess the fixed-node error.

We have chosen to study the diamond/$\beta$-tin transition in Si for
several reasons. First, it has been investigated by several
experimental groups~\cite{experiments}, with results that are
consistent enough for the present purpose. Second, there is already
considerable QMC experience with diamond-structure Si, from which it
is known that the cohesive energy is correct to within the
experimental error of $\pm$ 80 meV/atom, and the equilibrium lattice
parameter and bulk modulus are also accurately
reproduced~\cite{leung99,li91,kent99}. The third and most important reason
for studying this transition is that it is likely to be theoretically
troublesome, because of the significant change of electronic
structure. In the 4-fold coordinated diamond structure, Si is a
semiconductor, whereas in the 6-fold coordinated $\beta$-tin structure
it is a semi-metal. The difference in the electronic exchange and
correlation energies between the two phases is likely to lead to
non-cancelling errors. This is manifested in the serious
under-prediction of the transition pressure by the local density
approximation (LDA), the predicted value of
5.7~-~6.7~GPa~\cite{gaal-nagy99,boyer91,needs95,moll95,temperature_effect}
being only about half of the experimental value of 10.3~-~12.5
GPa~\cite{experiments} (in fact, a value of 8.8 GPa for the transition
pressure has also been reported by one experimental
group~\cite{olijnyk84}, but this is thought to be an
underestimate). This error is considerably reduced by the generalised
gradient approximation (GGA)~\cite{moll95}. Essentially the same DFT
errors lead to an LDA under-prediction of the Si melting temperature
by 23~\%~\cite{sugino95,alfe03}, reduced to 12~\% by the
GGA~\cite{alfe03}. The reason why such transitions are a sensitive
test of QMC (or any other total-energy method) is that rather small
changes in relative energies give substantial changes in the
transition pressure: in Si, an energy change of 100 meV/atom gives a
change in transition pressure of $\sim 3$~GPa.

The plan of the paper is as follows. In the next Section, we summarise
briefly the QMC techniques and describe in more detail the sources of
error and the measures we have taken to eliminate or reduce
them. Sec. 3 reports our numerical results, presenting first our
extensive tests on the different kinds of error and then our results
for the energies, volumes and transition pressure of the
diamond/$\beta$-tin transition and the comparison with experiment. In
Sec. 4, we discuss the implications of the work and draw conclusions.

\section{Methods} 

The VMC and DMC techniques used in this work have been described in
detail in reviews~\cite{foulkes01}, so here we recall rather briefly
the underlying ideas and outline the sources of error that we have
tried to bring under control.

The VMC method gives an upper bound on the exact ground-state energy
$E_0$. Given a normalised trial wavefunction $\Psi_T ( {\bf R} )$,
where ${\bf R} =({\bf r}_1, {\bf r}_2 \dots,{\bf r}_N)$ is a
$3N$-dimensional vector representing the positions of $N$ electrons,
and denoting by $\hat{H}$ the many-electron Hamiltonian, the
variational energy $E_v \equiv \langle \Psi_T | \hat{H} | \Psi_T
\rangle \ge E_0$ is estimated by sampling the value of the local
energy $E_L ( {\bf R} ) \equiv \Psi_T^{-1}({\bf R})\hat{H} \Psi_T({\bf
R})$ with configurations ${\bf R}$, distributed according to the
probability density $\Psi_T ({\bf R})^2$.  Our trial wavefunctions are
of the usual Slater-Jastrow type:
\begin{equation}
\Psi_T ( {\bf R} ) = D^\uparrow D^\downarrow e^{J} \; ,
\end{equation}  
where $D^\uparrow$ and $D^\downarrow$ are Slater determinants of up-
and down-spin single-electron orbitals, and $e^J$ is the so called
Jastrow factor, which is the exponential of a sum of one-body and
two-body terms, with the latter being a parametrised function of
electron separation, designed to satisfy the cusp conditions. The
parameters in the Jastrow factor are varied to minimise the variance
of the local energy $E_L$.  

In practice, VMC results are not accurate enough, and one needs to use
DMC. The basic idea is to compute the evolution of the many-body
wavefunction $\Phi$ by the time-dependent Schr\"{o}dinger equation in
imaginary time $- \partial \Phi\left/ \partial t \right. = (\hat{H} -
E_T) \Phi $, where $E_T$ is an energy offset. The equivalence of this
to a diffusion equation allows $\Phi$ to be regarded as a probability
distribution represented by a population of diffusing walkers. In
practice, it is essential to use ``importance sampling'', which means
computing the evolution of the function $f$ defined $f=\Phi \Psi_T$,
where the trial wavefunction $\Psi_T$ is a good approximation to the
true many-electron wavefunction, taken from the VMC calculations. The
time evolution of $f$ is given by:
\begin{equation}\label{eqn:diffusion}
-\frac{\partial f({\bf R},t)} {\partial t} = -\frac{1}{2} \nabla^2
f({\bf R},t) + \nabla \cdot [ {\bf v}_D({\bf R})f({\bf R},t)] +
[E_L({\bf R}) - E_T] f({\bf R},t),
\end{equation}
where ${\bf v}_D({\bf R}) \equiv \nabla \ln |\Psi_T({\bf R})|$ is the
$3N$-dimensional drift velocity and $E_L ( {\bf R} )$, as before, is
the local energy.  In principle, the DMC scheme yields the exact
ground state energy, but for fermion systems there is a fundamental
problem. This is that $\Phi$ changes sign as {\bf R} varies, so that
it can only be treated as a probability in regions of {\bf R}-space
where it does not change sign. These regions, and the nodal surfaces
that defines their boundaries, are necessarily those of the trial
wavefunction $\Psi_T$. The consequence is that the energy given by DMC
is not the true ground state energy but is an upper bound because of
the constraint that the nodal surface is that of $\Psi_T$. This gives
rise to the so called ``fixed-node'' error, which is one of the
concerns of this paper.

In summarising the technical questions that are important in this
work, we focus on the implementation of DMC, since this determines the
accuracy of the final results. The time evolution of the diffusing
walkers is computed using the Green's function technique in the
short-time approximation~\cite{foulkes01}. We shall present tests
showing that the time step used in this approximation can be chosen to
render errors negligible. For the representation of the
single-electron orbitals contained in the Slater determinants
$D^\uparrow$ and $D^\downarrow$, a number of basis sets have been used
in previous work, including plane-waves and Gaussians. In the present
work, we use a B-spline basis, also known as ``blip functions'',
consisting of piecewise continuous localised cubic spline functions
centred on the points of a regular grid. For a detailed account of
this basis set, and an explanation of the great advantages of using
this basis for QMC, see Ref.~\cite{alfe04}. The key point here is that
basis-set convergence is readily achieved simply by decreasing the
spacing $a$ of the blip grid. Roughly speaking, if the single-electron
orbitals would require a wavevector cut-off $k_{\rm max}$ for their
representation in a plane-wave basis, then the blip-grid spacing will
need to be $a \le \pi / k_{\rm max}$, and rapid convergence is
expected as $a$ is reduced below this value.

An important source of error in QMC calculations using periodic
boundary conditions is the limited size of the repeating cell.  In DFT
calculations one normally studies a primitive unit cell and integrates
quantities over the Brillouin zone, a procedure whose cost is
proportional to the number of $k$-points sampled.  This is equivalent
to studying a much larger unit cell with a single $k$-point.  In
many-body calculations, it is not possible to reduce the problem to one
within the primitive unit cell because the many-body Hamiltonian is
not invariant under the translation of a single electron by a
primitive lattice vector.  In other words, one has to use a large
simulation cell and solve at one $k$-point, and the cost is
proportional to the cube of the number of electrons in the cell.  This
means that converging QMC calculations with respect to system size is
much more costly than converging DFT ones.  We follow the common
practice~\cite{williamson98} of correcting for this error by using separate DFT
calculations: we add to the DMC energies the difference $\Delta
E_{\Gamma \rightarrow \bf k}$ between the DFT-LDA energy calculated
with a very large set of {\bf k}-points and the DFT-LDA energy
calculated using the same sampling as in the DMC calculation.

In this work, we used pseudopotentials generated by both Hartree-Fock
(HF) and LDA calculations on the Si atom. The non-locality that is
essential in these pseudopotentials gives rise to unavoidable errors
in DMC. The reason is that the diffusion equation with a non-local
Hamiltonian becomes:
\begin{equation}\label{eqn:diffusion_nonlocal}
-\frac{\partial f} { \partial t} = -\frac{1}{2} \nabla^2 f + \nabla
\cdot [ {\bf v}_D f ] + \frac{(\hat H - E_T)\Psi_T}{\Psi_T} f - \left
\{ \frac{ \hat V_{nl} \Psi_T}{\Psi_T} - \frac{ \hat V_{nl} \Phi}{\Phi}
\right \} f,
\end{equation}
where $\hat{V}_{nl}$ is the non-local component of the
pseudopotential. The last term in the equation can change its sign as
time evolves, and therefore presents the same difficulties as the
fermion sign problem. To avoid this difficulty one introduces the so
called ``localisation approximation'', in which the last term in
Eq.~\ref{eqn:diffusion_nonlocal} is simply neglected. If the trial
wavefunction $\Psi_T$ is close to the true (fixed node) ground state
wavefunction $\Psi$, then this approximation introduces an error which
is small and proportional to $(\Psi_T - \Psi)^2$~\cite{nonlocal}. This
error, however, is non-variational, so it can decrease as well as
increase the total energy.

We also tested the effect of adding a ``core polarisation potential''
(CPP)~\cite{maezono03} to the pseudopotential.  CPPs go beyond the
standard pseudopotential approximation, by describing the polarisation
of the atomic cores by the electrons and the other atomic cores.  In
the CPP approximation, the polarisation of a particular core is
determined by the electric field at the nucleus from the instantaneous
positions of the electrons and the other atomic cores.  CPPs therefore
account approximately for both dynamical core-valence correlation
effects and static polarisation effects.  Our implementation of CPPs
within QMC calculations is described in Ref.~\cite{maezono03}, and we
used the CPP parameters reported in Ref.~\cite{shirley93}.

Details of the {\sc casino} code used in all the QMC calculations
are given in Ref.~\cite{needs04}. In order to suppress
statistical bias in the total energy, QMC calculations
need to be run with a large population of walkers,
and this makes it efficient to run on massively parallel
machines, with parallelism achieved by distributing
walkers across processors. 

\section{Results} 

\subsection{Tests} 

We present here the results of our tests on error sources; results on
the diamond/$\beta$-tin transition itself are reported in Sec.~3.2. To
provide a framework for the discussion of errors, we set ourselves the
target of reducing the sum of all controllable errors below
30~meV/atom, this value being chosen because it corresponds to an
error $\sim 1$~GPa in the transition pressure. The sources of
controllable error are: sampling statistics, time step, blip-grid
spacing, cell size, pseudopotentials, localisation approximation, and
CPP.

\begin{figure}
\psfig{figure=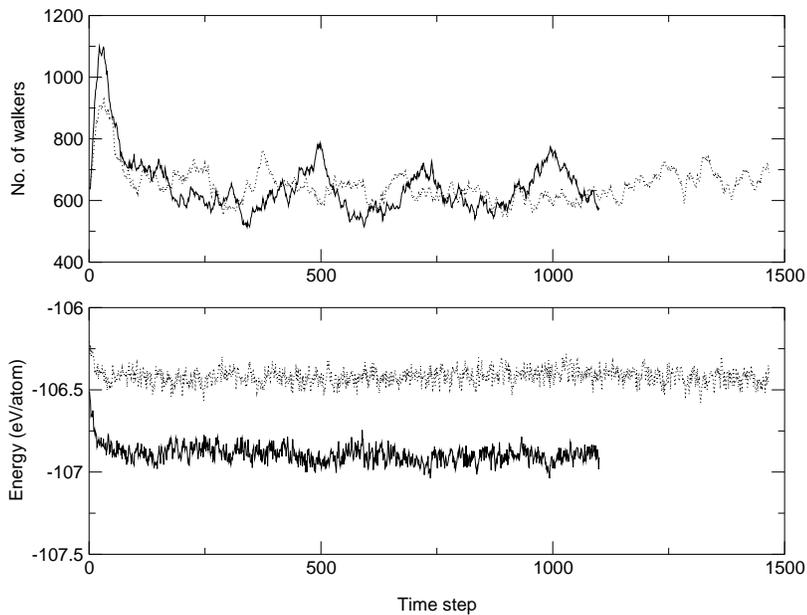,height=3.4in,angle=-90}
\caption{Lower panel: DMC local energies as function of time
(time-step=0.03 a.u.) for the $\beta$-tin structure with
$V=15$~\AA$^3$/atom (dotted line) and the diamond structure with
$V=20$~\AA$^3$/atom (continuous line). Upper panel: the population of
walkers for the $\beta$-tin structure (dotted line) and the diamond
structure (continuous line).}\label{fig:dmc_runs}
\end{figure}

To ensure that sampling bias is negligible, our DMC calculations are
run with a target population of 640 walkers for both crystal
structures. With this number of walkers, the statistical error
necessarily falls well below our threshold of 30~meV/atom. The reason
for this is that the DMC decay to the ground state occurs after $\sim
100$ steps, but the calculations need to extend over $\sim 1000$ steps
to ensure complete stability. With the cell sizes used in this work
and the number of walkers we employ, the statistical error after $\sim
1000$ steps is already less than 5~meV/atom. As an illustration of
this, Fig.~\ref{fig:dmc_runs} shows results from typical simulations
of the diamond and $\beta$-tin structures, close to their equilibrium
volumes. The rapid decay to the ground state is clear, and one also
notes the stability of the walker population around the target value
of 640. For this number of walkers, the DMC calculations are efficient
on up to 128 processors. Beyond this processor number, parallel
scaling worsens, because fluctuations in the number of walkers start
to cause inefficient load balancing.

\begin{figure}
\psfig{figure=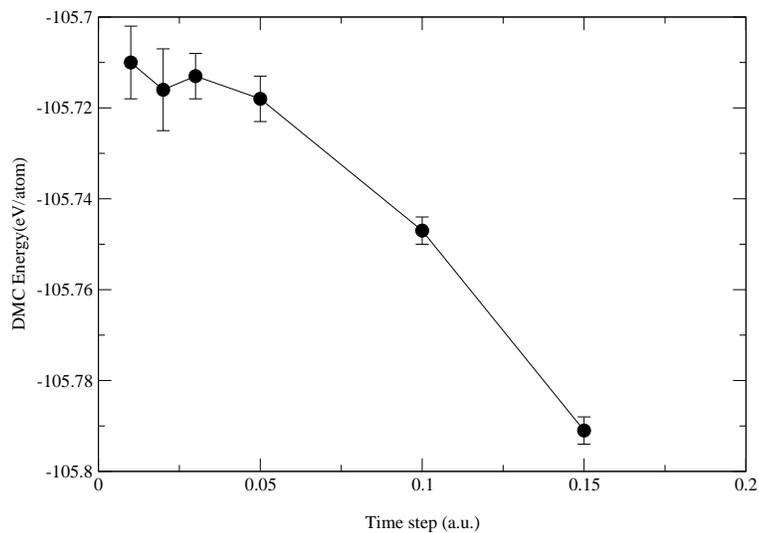,height=3.4in,angle=-90}
\caption{The DMC total energy per atom as a function of time step,
with error bars showing the statistical errors. The calculations were
performed using a cell containing 16 atoms in the $\beta$-tin
structure at the volume $V=15$~\AA$^3$/atom.}\label{fig:time_step}
\end{figure}

In Fig.~\ref{fig:time_step} we show tests on time step errors,
performed with a cell containing 16 atoms in the $\beta$-tin structure
at the volume $V = 15$~\AA$^3$/atom, which is close to the calculated
DFT-LDA equilibrium volume.  We tested time steps between 0.01
a.u. and 0.15 a.u., with the length of the runs chosen so that the
statistical error was less than 10~meV/atom. The results show that
with a time step of 0.03 a.u. the error is smaller than the target
accuracy, and we therefore used a time step of 0.03 a.u. for our final
calculations.

The blip-function basis set~\cite{alfe04} was also tested with the
16-atom $\beta$-tin cell and $V = 15$~\AA$^3$/atom. In
Table~\ref{tab:blip_tests} we report the values of the kinetic energy,
the local potential energy and the non-local potential energy
calculated using DFT, VMC and DMC, both with plane-waves (PW)
and blips. The PW results were obtained using the
{\sc pwscf} code~\cite{pwscf} with a PW cutoff energy of 15 Ry.  For the
purpose of these tests, we did not use a Jastrow factor in the VMC
calculations, so that the three energy terms should have exactly the
same values in DFT and VMC.  This is clearly the case for the VMC
performed using PW, but there is a difference of up to 60~meV/atom
when the VMC calculations are performed using the blip representation
with the natural grid ($a=\pi/k_{\rm max}$). However, this difference
is reduced to less than 5~meV/atom if a grid of half the spacing is
used ($a=\pi/ 2 k_{\rm max}$). This proves that, provided the blip grid
is dense enough, the results are indistinguishable from those obtained
using plane waves. For DMC calculations, a Jastrow factor has been
included in the trial wavefunction, and the situation is more
interesting. Even though the natural grid does not provide a perfect
description of the single-particle orbitals, the DMC total energy is
essentially the same as that calculated with PW. Of course, this is
what one would expect in a perfect DMC calculation, since the total
energy is independent of the trial wavefunction. However, with the
fixed-node and pseudopotential localisation approximations, the total
energy does in general show a weak dependence on the trial
wavefunction.

Tests on the size of the simulation cell were performed on both the
diamond and the $\beta$-tin structures, with cells containing up to
432 atoms. Since we were mainly interested in the
diamond~$\rightarrow$~$\beta$-tin transition pressure, we calculated
the energies at the two volumes $V = 20$~\AA$^3$/atom and $V =
15$~\AA$^3$/atom for the diamond and $\beta$-tin structures,
respectively, which are both close to the calculated equilibrium
volumes. As far as the transition pressure is concerned, the important
quantity to test is the energy difference between the phases at the
two volumes.  The results of the tests are reported in
Table~\ref{tab:size}, where we also report the values of the energies
extrapolated to infinite cell size $E_{\rm tot}^\infty$ for the two
structures. These are obtained by linear extrapolation to $1/N$, with
$N$ the number of atoms in the repeating cell, between the {\bf
k}-points corrected results obtained with the two cells containing 128
and 432 atoms.  The cell size errors obtained with 128-atom cells are
about 110 meV/atom, and they are approximately the same in the two
structures.  This indicates that the residual size error can be
regarded as a constant energy offset, which will not affect physical
properties such as structural parameters and the
diamond~$\rightarrow$~$\beta$-tin transition pressure, and we
therefore chose to use cells containing 128 atoms.

We tested both a HF and a LDA pseudopotential. In both cases, the local
part of the pseudopotential was chosen to be the $p$ angular momentum
component. The tests were performed once again on the two structures
with $V = 15$~\AA$^3$/atom and $V = 20$~\AA$^3$/atom for the
$\beta$-tin and the diamond structures, respectively. We found that
the energy differences between the two structures were 0.535(5) and
0.550(5) eV/atom for the HF and the LDA pseudopotentials
respectively. The two numbers are very close, which indicates that the
choice of the pseudopotential does not affect the results
significantly. However, we believe that in a QMC calculation it is
more consistent to use a HF pseudopotential rather than an LDA one,
because the former does not build in any correlation. We therefore
used the HF pseudopotential.

To test the pseudpotential localisation approximation, we performed
additional calculations with the HF pseudopotential by changing the
local part of the pseudopotential to the $s$ angular momentum
component. Calculations were performed on the diamond structure with a
16-atom cell and $V=20$~\AA$^3$/atom. Clearly, by changing the local
part of the pseudopotential there is no guarantee that the quality of
the pseudopotential does not change, therefore it is conceivable that
the total energy may change simply because the pseudopotential has
changed. So we have first performed a DFT-LDA calculation with this
pseudopotential, and found a difference of less than 2 meV/atom when
the local part is changed from $p$ to $s$, which is extremely small
for our purposes. We then performed a DMC simulation with the HF
pseudopotential having the $s$ channel as the local part, and within a
statistical error of 10~meV/atom we found no energy difference from
the calculation with the HF pseudopotential and the $p$ channel as the
local part. This indicates that the error from the localisation
approximation is probably less than 10~meV/atom in this case.

As a final test on the pseudopotential, we considered the inclusion of
a CPP. We expect the CPP energy to be more important in the
$\beta$-tin structure, which has a smaller volume and therefore the
electrons and ions are on average closer to one another.  Tests were
performed on both structures at the two volumes $V = 15$~\AA$^3$/atom
and $V = 20$~\AA$^3$/atom for the $\beta$-tin and the diamond
structures, respectively. We found that with the CPP the energy
difference between the two calculations was 0.505 (10) eV/atom, which
is slightly lower than the value of 0.535 (10) eV/atom obtained
without the CPP. The inclusion of this correction has a small effect
on the transition pressure which will be discussed in the next
section.

\subsection{Results}

\begin{figure}
\psfig{figure=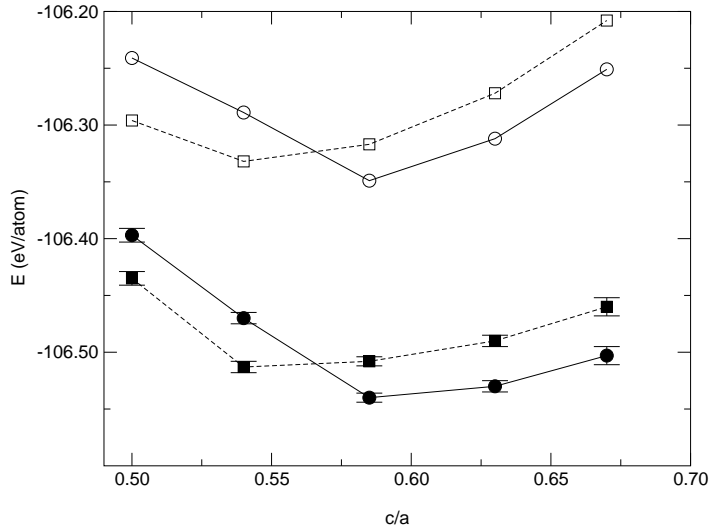,height=3.4in,angle=-90}
\caption{DMC energies for the $\beta$-tin structure at the volume
$V=15$~\AA$^3$/atom as a function of the $c/a$ ratio, performed with
cells containing 128 atoms (filled circles). DFT results performed
with the equivalent mesh of {\bf k}-points are also shown (open
circles). DMC {\bf k}-points corrected results are shown as filled
squares, and DFT results fully converged with respect to {\bf k}-point
sampling (open squares). The lines are guides to the eye.}\label{fig:c_su_a}
\end{figure}

We now turn to the energetics of the diamond and $\beta$-tin
structures and the transition pressure between them.  Since the
$\beta$-tin structure is body-centred tetragonal, its energy depends
not only on volume, but also on the $c/a$ ratio.  For each volume, we
should therefore minimise the energy with respect to the $c/a$
ratio. However, using DFT-LDA calculations we found that the minimum
of the energy depends rather weakly on $c/a$, and that choosing
$c/a=0.54$ for all volumes of interest only affects the energy by a
few meV/atom. To check that this $c/a$ ratio is also appropriate within
DMC, we have performed DMC calculations at five different
$c/a$ ratios for $V=15$~\AA$^3$/atom. The results of the test are
displayed in Fig.~\ref{fig:c_su_a}, where we report the DMC raw data
and the $k$-points corrected results. By interpolating the DMC data, we
find that the DMC minimum is at $c/a=0.554$, which is very close to
the experimental value $c/a=0.552$. For comparison, we also report
calculations for the same structures performed with DFT-LDA. It is
clear that the dependence of the energy on the $c/a$ ratio is very
similar in the two techniques, and therefore we chose to use
$c/a=0.54$ for all calculations.

\begin{figure}
\psfig{figure=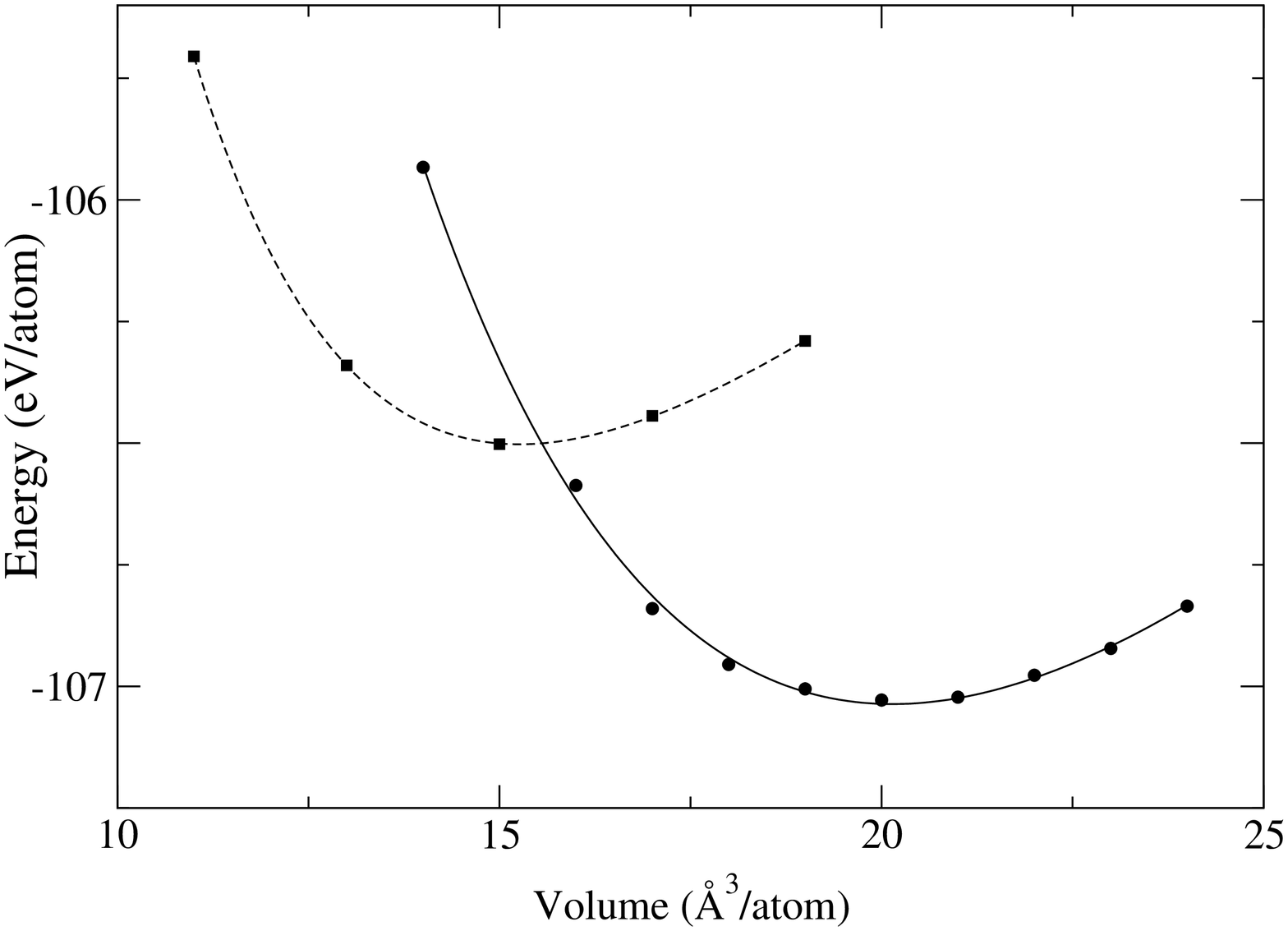,height=3.4in,angle=-90}
\caption{DMC total energies for the $\beta$-tin (squares) and the
diamond (circles) structures. The size of the points corresponds to
about two standard deviations. The dashed and continuous lines are
Birch-Murnaghan EOS curves fitted to the data.}\label{fig:e_of_v}
\end{figure}

In Fig.~\ref{fig:e_of_v} we report the calculated energies $E(V)$ for
the two structures corrected for {\bf k}-points errors. CPP
corrections are not included in these results.  These energy points
were then used to fit the parameters of the Birch-Murnaghan equation
of state:
\begin{equation}\label{murna}
E = E_0 + \frac{3}{2}V_0B_0 \left [ \frac{3}{4}(1+2\xi)\left
(\frac{V_0}{V}\right )^{4/3} - \frac{\xi}{2} \left ( \frac{V_0}{V}
\right )^{2}
-\frac{3}{2}(1+\xi) \left ( \frac{V_0}{V} \right )^{2/3} + \frac{1}{2}
\left ( \xi + \frac{3}{2}\right ) \right ] \hspace{12cm} \; , 
\end{equation} 
where $\xi = (3 - 3 B_0'/4)$, $V_0$ is the equilibrium volume, $B_0$
the zero-pressure bulk modulus, $B_0'$ its derivative with respect to
pressure at zero pressure, and $E_0$ the energy minimum.  The fitted
curves are also reported on the same Figure. The values of the fitted
parameters are reported in Table~\ref{tab:murna_parameters} together
with DFT-LDA and DFT-GGA results and experimental data. We also report
in the Table previous DMC results obtained by Li \textit{et
al}.~\cite{li91} for the diamond structure.
The agreement with the experimental data is extremely good, and is
also somewhat better than obtained previously by Li \textit{et
al}~\cite{li91}. In particular, the equilibrium volume is overestimated
by only 0.5 \%. In comparison, DFT-LDA underestimates the equilibrium
volume by 2 \%, and the two DFT-GGA BP and PW91~\cite{moll95}
overestimate it by 1\% and 2\%, respectively.

Using our results, we obtain a DMC transition pressure of $\sim
19$~GPa. Before comparing our calculated transition pressure with the
experiments, we note that our calculations do not include zero-point
motion, which has been shown to be different in the two
phases. Moreover, experimental transition pressures are only reported
at room temperature, therefore there is a significant contribution to
the free energy coming from the difference in vibrational free
energies between the two structures. As shown by Ga\'al-Nagy
\textit{et al}.~\cite{gaal-nagy99}, the zero point motion stabilises
the $\beta$-tin structure with respect to the diamond structure, and
lowers the transition pressure by about 0.3 GPa. At room temperature
the stabilisation of the $\beta$-tin structure lowers the transition
pressure by an additional 1 GPa, so that the two effects lower the
transition pressure by $\sim$~1.3 GPa. If we add this correction to
our calculated transition pressure we obtain 17.7 GPa.  Moreover,
these calculations did not include CPP corrections, which reduces the
free energy of the $\beta$-tin structure relative to the diamond
structure. If we assume that these corrections are approximately the
same at different volumes in the two structure, then we finally obtain
a corrected transition pressure of 16.5 GPa. The experimental
transition pressure is in the range 10.3~-~12.5 GPa~\cite{experiments},
which is significantly lower than predicted by our calculations.

\section{Discussion} 

We recall that the main purpose of this work is to assess the accuracy
of QMC for Si by examining its prediction for the transition pressure
between the diamond and $\beta$-tin structures. However, we discuss
first the controllable sources of error that we have attempted to
reduce below our threshold of 30~meV/atom.

We have shown that errors due to statistical sampling and finite time
step are readily reduced to negligible size.  Convergence with respect
to basis set completeness is also easy to achieve, and we have noted
the important advantages of the B-spline (blip) basis, which combines
ease of convergence with excellent scaling with respect to system
size, as reported in detail elsewhere~\cite{alfe04}.  System size
errors also appear to be under excellent control. By performing DMC on
cells of up to 432 atoms, we have shown that the error in total energy
is reduced to $\sim 110$~meV/atom, but the size error on the
difference in energy between the diamond and $\beta$-tin structures is
reduced to less than $\sim 5$~meV/atom. Errors due to the
pseudopotential approximation itself, as well as to the
pseudopotential localisation approximation, also appear to be no
larger than $\sim 5$~meV/atom, though we have not shown this
rigorously. Finally, we have studied the effect of including core
polarisation, and shown that this reduces the energy difference
between the two structures by $\sim 30$~meV/atom. Taken together,
these tests suggest that if there were no other sources of error, the
transition pressure could be calculated to within $\sim 1$~GPa.

Our results for diamond-Si confirm the excellent accuracy of DMC for
this structure. Our cohesive energy agrees with the experimental value
within the experimental error of $\sim 80$~meV/atom, the equilibrium
lattice constant is correct to 0.2~\% and the bulk modulus to
3~\%. However, for the transition pressure, our DMC result of 16.5~GPa
is significantly larger than the experimental range of
10.3~-~12.5~GPa. This less than satisfactory agreement could in
principle be due either to uncertainty in the experimental results or
to remaining errors in the QMC calculations. We think it unlikely that
experiments could underestimate the equilibrium transition pressure by
such a large amount. There appears to be a large barrier to the
transition on increase of pressure, and the transition is in fact
irreversible, with complex tetrahedral phases being formed on release
of pressure~\cite{experiments,mujica03}.  If anything, this
irreversibility would make it more likely for the experimental values
to be too high. On the theoretical side, we have shown that most of
the sources of error are too small to account for the discrepancy. The
only remaining theoretical error that could be large enough is the
fixed-node error.  Since the fixed node error can only increase the
energy, and since the DMC transition pressure is too high, a possible
scenario is that the fixed-node error raises the energy of
$\beta$-tin-Si relative to diamond-Si.

In conclusion, we have shown the feasibility of using QMC to calculate
the relative stability of different crystal structures, with most
technical errors reduced enough to give the transition pressure to
within $\sim 1$~GPa. Nevertheless, the computed transition pressure
for the diamond $\rightarrow$ $\beta$-tin transition in Si differs
from the experimental value by $\sim 4-6$~GPa. The evidence presented
indicates that the discrepancy may be due to QMC fixed-node error.

\section*{Acknowledgments}
The computations were performed on the CSAR and the HPCx services,
using allocations of time from NERC through the Mineral Physics
Consortium and from EPSRC through the UKCP Consortium. Calculations
have also been performed on the Altix machine at University College
London provided by the SRIF programme.  DA acknowledges support from the
Royal Society and the Leverhulme Trust. We gratefully acknowledge
helpful discussions with Cyrus Umrigar.

\begin{table}
\begin{tabular}{lcccc}
\hline
  &  & PW & Blips ($a=\pi/k_{\rm max}$) & Blips ($a=\pi/2k_{\rm max}$) \\ 
\hline
DFT & & & & \\
& $E_{\rm kin}$ & 43.863 & & \\
& $E_{\rm loc}$ & 15.057 & & \\
& $E_{\rm nl}$  & 1.543 & & \\
 & & & & \\
VMC & & & & \\
& $E_{\rm kin}$ & 43.864(3) & 43.924(3) & 43.862(3) \\
& $E_{\rm loc}$ & 15.057(3) & 15.063(3) & 15.058(3) \\
& $E_{\rm nl}$  & 1.533(3)  & 1.525(3)  & 1.535(3) \\
& $E_{\rm tot}$  & $-101.335(3)$ & $-101.277(3)$ & $-101.341(3)$ \\
& & & & \\
DMC & & & & \\
& $E_{\rm tot}$  & $-105.714(4)$ & $-105.713(5)$ & $-105.716(5)$ \\
\hline
\end{tabular}
\caption{Total energy $E_{\rm tot}$ and the kinetic energy, local
pseudopotential energy and non-local pseudopotential energy components
$E_{\rm kin}$, $E_{\rm loc}$ and $E_{\rm nl}$ calculated using plane
wave (PW) basis sets and blip function basis sets with two grid
spacings $a$ (energy units: eV/atom). Results are from DFT, VMC and
DMC calculations on Si in the $\beta$-tin structure, with a repeating
cell of 16 atoms. A Jastrow factor was included only in the DMC
calculations. Blip-grid spacing $a$ is specified in terms of the PW
cut-off wavevector $k_{\rm max}$, corresponding to a cut-off energy of
15 Ry.}\label{tab:blip_tests}
\end{table}

\begin{table}
\begin{tabular}{lcccc|cccc|c}
\hline
         &   \multicolumn{4}{c}{128}     &    \multicolumn{4}{c}{432}  & \\
    & $E_{\Gamma}$ & $\Delta E_{\Gamma \rightarrow \bf k}$ & $E_{\rm tot}$ & $\delta E$ &$E_{\Gamma}$ & $\Delta E_{\Gamma \rightarrow \bf k}$ &  $E_{\rm tot}$ & $\delta E$ &  $ E_{\rm tot}^\infty$ \\
\hline
diamond & -106.926(5) & -0.102  & -107.028(5) & 0.108(5) & -106.937(6) & -0.014 & -106.951(6) & 0.031(6) & -106.920(5)\\
$\beta$-tin & -106.457(5) & -0.045 & -106.502(5) & 0.117(5) & -106.416(7) & -0.004 & -106.420(7) & 0.035(7) & -106.385(5) \\
\hline
\end{tabular}
\caption{DMC total energies (eV/atom) for the diamond and the
$\beta$-tin structures of Si, obtained from calculations on 128-atom
and 432-atom cells.  $E_{\Gamma}$ is the raw DMC energy, $\Delta
E_{\Gamma \rightarrow \bf k}$ is the {\bf k}-points correction,
$E_{\rm tot}$ is the sum of $E_{\Gamma}$ and $\Delta E_{\Gamma
\rightarrow \bf k}$, and $E_{\rm tot}^\infty$ is the value of $E_{\rm
tot}$ extrapolated to infinite cell size. For each cell size, $\delta
E$ is the difference $E_{\rm tot} - E_{\rm
tot}^\infty$.}\label{tab:size}
\end{table}

\begin{table}
\begin{tabular}{lcccccc}
\hline
  & Expt. & LDA & BP & PW91 & DMC & DMC (This work) \\
\hline
Diamond  & & & & & \\
  $V_0$~(\AA$^3$) & 20.01$^a$ & 19.57$^b$ & 20.46$^b$ & 20.23$^b$ & 20.23 (20)$^c$ & 20.11 (3) \\
  $B_0$~(GPa)        & 99$^a$  & 97$^b$  & 90$^b$  & 92$^b$  & 103 (7)$^c$ & 103 (10)  \\
  $E_{coh}$~(eV)  &  4.62(8)$^d$ & 5.338$^e$ &  & 4.653$^e$  & 4.51(3)$^c$, 4.63(2)$^f$ & 4.62(1) \\
  & & & & & & \\
$\beta$-tin  & & & & & \\
  $V_0$~(\AA$^3$) &  & 14.63$^b$ & 15.84$^b$ & 15.67$^b$ & &  15.26 (3) \\
  $B_0$~(GPa)         &  & 115$^b$   & 99$^b$    & 104$^b$   & & 114 (5)  \\
  $E_{coh}$~(eV)      &      & 5.115$^e$ &  & 4.313$^e$ &  & 4.10(1) \\
  $c/a$              & 0.552$^a$ & 0.548$^b$ &   & 0.548$^b$  & & 0.554 \\
\hline
  $\Delta E_0$ (eV) &   & 0.226$^b$ & 0.404$^b$ & 0.341$^b$ & & 0.505 (10) \\ 
  $p_t$ (GPa) & $10.3-12.5^g$ & 6.7$^b$ & 13.3$^b$ & 10.9$^b$ & & 16.5 (5) \\
\hline
\end{tabular}
$^a$Cited in Ref.~\protect\cite{yin82}\\
$^b$Ref.~\protect\cite{moll95}\\
$^c$Ref.~\protect\cite{li91} \\
$^d$Ref.~\protect\cite{janaf85}\\ 
$^e$Ref.~\protect\cite{hennig04}\\
$^f$Ref.~\protect\cite{leung99}\\
$^g$Ref.~\protect\cite{experiments}
\caption{Structural properties and the diamond~$\rightarrow \beta$-tin
transition pressure $p_t$, calculated within DFT using different
exchange-correlation functionals and within DMC. $V_0$ is the
equilibrium volume, $B_0$ the zero pressure bulk modulus, $E_{coh}$ is
cohesive energy, $\Delta E_0$ is the calculated difference of minimum
energy between the two structures, and $c/a$ is the ratio of
tetragonal lattice parameters of the $\beta$-tin structure.
Theoretical cohesive energies have all been corrected for zero point
motion (0.06 eV and 0.04 eV in the diamond and $\beta$-tin structures,
respectively). The structural parameters calculated within DMC for the
diamond structure were obtained from a Birch-Murnaghan equation of
state fit to energies calculated at volumes between 17 and
24~\AA$^3$/atom.  For the $\beta$-tin structure volumes between 11 and
19\AA$^3$/atom were used.  The experimental data are at 0 K and 77 K,
for the zero pressure equilibrium volume and bulk-modulus,
respectively.  The transition pressures have been corrected for finite
temperature effects evaluated at 300 K~\protect\cite{gaal-nagy99}. The
DMC calculations for the transition pressure also include a
CPP. }\label{tab:murna_parameters}
\end{table}

\begin{thebibliography}{99}
\bibitem{temperature_effect} The range 5.7~-~6.7 GPa for the LDA
transition pressure of Si from the diamond to the $\beta$-tin
structure also includes the effect of the phonon free energy at room
temperature~\cite{gaal-nagy99}, which lowers the transition pressure
by about 1.3 GPa with respect to the transition pressure calculated
without including this effect.
\end{thebibliography}
\end{document}